\documentclass[aps,pre,preprint,groupedaddress,eqsecnum,amssymb]{revtex4}

\usepackage{graphicx}

\begin{document}


\title{Calculation of the persistence length of a flexible polymer
  chain with short range self--repulsion}


\author{Lothar Sch\"afer}
\email[E--mail: ]{lsphy@theo-phys.uni-essen.de}
\author{Knut Elsner}
\affiliation{Universit\"at Duisburg-Essen
  -- Standort Essen --\\
  Universit\"atsstr.\ 5, 45117 Essen, Germany}


\date{July 14, 2003}

\begin{abstract}
  For a self--repelling polymer chain consisting of $n$ segments we
  calculate the persistence length $L\left(j,n\right)$, defined as the
  projection of the end--to--end vector on the direction of the
  $j^\textrm{\scriptsize th}$ segment. This quantity shows some pronounced
  variation along the chain. Using the renormalization group and
  $\epsilon$--expansion we establish the scaling form and calculate
  the scaling function to order $\epsilon^2$. Asymptotically the
  simple result $L\left(j,n\right) \approx \mbox{const}
  \left(j\left(n-j\right)/n\right)^{2\nu-1}$ emerges for dimension $d
  = 3$. Also outside the excluded volume limit $L\left(j,n\right)$ is
  found to behave very similar to the swelling factor of a chain of
  length $j \left(n-j\right)/n$. We carry through simulations which
  are found to be in good accord with our analytical results. For $d =
  2$ both our and previous simulations as well as theoretical
  arguments suggest the existence of logarithmic anomalies.
\end{abstract}


\maketitle

\section{\label{sec:1}Introduction}

In solution a long flexible polymer chain takes a random, coil--like
configuration. Many properties of these coils adequately are described
by a simple model, where the chain is taken as a linear sequence of
$n$ structureless segments ${\bf s}_{j} = {\bf r}_{j} - {\bf
  r}_{j-1}$, with vectors ${\bf r}_{j}$ ($j=0, \dots , n $), giving
the positions of the endpoints of the segments in $d$--dimensional
space. In such models the chemical microstructure of the polymer is
absorbed into a few parameters like the mean squared segment size
$\ell^2 \approx \langle {\bf s}^2_j \rangle$ or the excluded volume
$u_0$, which measures the interaction among the segments. For $u_0 >
0$ long chains are swollen compared to a noninteracting random flight
chain, and it is this `excluded volume region', which will be
considered here. We will discuss the influence of the excluded volume
on the persistence length, which measures the range over which the
chain configuration on average remembers the direction of a specific
segment. It is defined \cite{S1} as the projection of the end--to--end
vector ${\bf r}_n - {\bf r}_0$ on segment vector ${\bf s}_j$:
\begin{equation}
L(j,n) = \frac{\langle {\bf s}_j \cdot \left( {\bf r}_n - {\bf
      r}_0\right) \rangle}{\sqrt{\langle {\bf s}^2_j \rangle}} \: \: \: .
\end{equation}
Here the pointed brackets denote the thermodynamic average. We note
that often the persistence length is identified with $L(1,n)$, i.e.
with the projection on the first segment. We here use the more general
definition (1.1), since the position along the chain of the
distinguished segment ${\bf s}_j$ will turn out to be an essential
variable. Indeed, $L(j,n)$, varies considerably along the chain. Close to the chain ends ($j/n \to 0$ or 1, respectively) it is a nonuniversal, microscopic quantity, whereas in the critical limit ($n \to \infty$, $0 < j/n < 1$, fixed) it becomes a universal macroscopic observable of the excluded volume coil.

A microscopic, nonuniversal persistence length is an important parameter of the model of a `worm like' chain \cite{S1}, where bond--angle constraints correlate the
directions of subsequent segments, but no excluded volume interactions
among segments spaced a large distance along the chain exist. In that
model the persistence length is a measure of the local stiffness of
the chain and asymptotically becomes independent of the chain length
$n$.  The concept of a persistence length also plays a prominent role
in theories of polyelectrolyte solutions. Here the persistence length
refers to an underlying worm--like chain model, and various
definitions including also $L(j,n)$, Eq.~(1.1), are used. (See Refs.~\cite{S2,S3,S3b}
for recent work and a compilation of literature relevant to this
topic.) All this work treats the persistence length as a more local
quantity, independent of chain length. Excluded volume effects are
neglected.

If we take the excluded volume into account, we easily see that the
persistence length, as defined in Eq.~(1.1), cannot be independent of
chain length. Rather it must show power law behavior as function of
$n$. This follows from a sum rule, relating $L(j,n)$ to the
mean--squared end--to--end distance
\begin{equation}
R^2_e (n) = \langle \left( {\bf r}_n - {\bf r}_0 \right)^2 \rangle \: \: \: .
\end{equation}
Taking $\langle {\bf s}^2_j\rangle = \textrm{const}$, which
should be a very good approximation, we immediately find
\begin{equation}
\sum^{n}_{j=1} L(j,n) \sim R^2_e(n) \sim n^{2\nu} \: \: \: ,
\end{equation}
implying
\begin{equation}
L(j,n) \sim n^{2 \nu -1} \: \: \: .
\end{equation}
This result should be valid in the excluded volume limit of long
self--repelling chains. Since $\nu > 1/2$ ($\nu \approx 0.588$
for $d=3$), this shows that $L(j,n)$ is a critical quantity, diverging
with increasing chain length. It thus is somewhat surprising that it
has not found much attention among workers concerned with the excluded
volume problem. We are aware of only three papers \cite{S4,S5,S6},
where $L(j=1,n)$ is calculated for self avoiding lattice walks, using
exact enumeration or Monte Carlo methods. Most results given there are
for two--dimensional lattices, where the results suggest that $L(1,n)$
diverges logarithmically \cite{S5,S6} or with a very small power
\cite{S4} of $n$. On the cubic lattice $L(1,n)$ seems to tend to a
constant \cite{S6}. In view of the sum rule (1.3) these results
immediately imply that $L(j,n)$ must be strongly dependent on $j$, as stressed above.

In the present work we use renormalized perturbation theory to
calculate $L(j,n)$ for a chain with short range self--repulsion. We  concentrate on the dependence on segment index $j$, where
in three dimensions we will find a surprisingly simple asymptotic
behavior. In two dimensions, however, the behavior of $L(j,n)$ is more complicated, and our results support the existence of logarithmic anomalies. For reasons explained later we in our calculation go to
second order in the renormalized coupling constant (two loop). We
furthermore compare our results to simulation data for chains up to
length $n = 2000$ on a cubic lattice. Simulation results for a square
lattice are also given. Our analysis is an extension of our previous
work \cite{S7} on the correlation function $\langle {\bf s}_{j_1}
\cdot {\bf s}_{j_2}\rangle$ of individual segment directions.

This paper is organized as follows. In Sect.~\ref{sec:2} we define the
model and introduce the general structure of perturbation theory. In
Sect.~\ref{sec:3} we evaluate $L(j,n)$ to two loop order. The one loop
result is analyzed with the help of a crossover formalism suggested by
a blob picture.  Such an approach is known \cite{S8} to yield good
results for a large variety of observables for dilute or semidilute
polymer solutions. The two loop calculation, evaluated in strict
$\epsilon$--expansion $(\epsilon = 4 - d)$, serves to support the
assumptions implicit in the crossover model. In Sect.~\ref{sec:4} we
compare our results to our simulations in three dimensions and also
discuss the logarithmic anomalies showing up in $d=2$.
Sect.~\ref{sec:5} summarizes our results.

\section{\label{sec:2}The model}
Our model has been presented in detail in Refs.~\cite{S7,S8}, and we
here briefly recall the essential features.

As mentioned in the introduction, we describe the chain configuration
by the set of vectors ${\bf r}_j$ $(j=0, \dots, n)$. The energy is
written as
\begin{equation}
\frac{H}{k_B T} = {\cal V}_0 + {\cal V}_2 \: \: \: ,
\end{equation}
where
\begin{equation}
{\cal V}_0 = \sum^{n}_{j=1} \frac{\left( {\bf r}_j - {\bf r}_{j-1}
  \right)^2}{4 \ell^2_0}
\end{equation}
incorporates the connectivity of the chain, and
\begin{equation}
e^{- {\cal V}_2} = {\prod_{j < j'}}^{'} \left[ 1 - \left(4 \pi \ell^2_0\right)^{d/2} \beta_e \delta^d \left( {\bf r}_j - {\bf r}_{j'}
 \right)  \right]  \end{equation}
represents the excluded volume interaction among the segments. Here
the excluded volume is written as $u_0 = \left(4 \pi
  \ell^2_0\right)^{d/2} \beta_e$, introducing the dimensionless
excluded volume parameter $\beta_e$. The microscopic length scale
$\ell_0$ governs the segment size. 
In relation to more microscopic modells it incorporates all
microstructure effects usually adressed as intrinsic stiffnes of the
chain. For $\beta_e = 0$ one finds
\begin{equation}
\langle {\bf s}^2_j \rangle_0 = 2 d \ell^2_0 \: \: \: .
\end{equation}
In Eq.~(2.3) the prime at the product indicates that in multiplying
out we omit all terms in which some vector ${\bf r}_j$ occurs more
than once.

To calculate $L\left(j,n\right)$ we define a generating functional
\begin{equation}
{\cal Z} \left({\bf q},{\bf h}\right) = \frac{\left(4 \pi \ell^2_0\right)^{d/2}}{\Omega} \int\limits_\Omega {\cal D} [\mathbf{r}] \,e^{- {\cal V}_0 -{\cal V}_2} e^{{\bf h} \cdot {\bf s}_j + i {\bf q} \left({\bf r}_n - {\bf r}_0\right)} \: \: \: ,
\end{equation}
\begin{equation}
{\cal D} \left[\mathbf{r}\right] = \prod^{n}_{j=0} \frac{d^d r_j}{\left(4 \pi \ell^2_0\right)^{d/2}} \: \: \: ,
\end{equation}
where $\Omega$ denotes the volume of the system. The normalizing
factors are chosen such that the partition function
\begin{equation}
Z \left(n\right) = {\cal Z} \left(0,0\right)
\end{equation}
reduces to $Z(n) = 1$ for a noninteracting system $\left(\beta_e = 0
\right)$. From ${\cal Z}\left({\bf q},{\bf h}\right)$ we can calculate
the (not normalized) persistence length as
\begin{equation}
\hat{L}\left(j,n\right) = \langle {\bf s}_j \cdot \left({\bf r}_n -{\bf r}_0 \right) \rangle = \frac{1}{Z\left(n\right)} \left(- i\,\nabla_q \cdot \nabla_h\right) {\cal Z} \left({\bf q}, {\bf h} \right) \bigg|_{{\bf q}\, = \,\mathbf{0}\, =\, \mathbf{h}} \: \: \: .
\end{equation}

The generating functional (2.5) is evaluated by expanding the product
(2.3) in powers of $\beta_e$. The individual contributions can be
represented by diagrams in which the polymer is drawn as a straight
line and the excluded volume interaction is represented by broken
lines (`vertices'), connecting pairs of special points which
correspond to the interacting segments ($j_\alpha$, $j_\beta$) (see,
e.g. Fig.~\ref{fig:2}). Parts of the polymer line connecting two consecutive
special points will be addressed as propagator lines, labeled by some
`momentum' variable ${\bf k}$.  Momentum conservation holds for each
vertex. The chain ends $0$, $n$ represent special points closing
propagator lines of momentum ${\bf q}$ or $-{\bf q}$, respectively.

In evaluating a diagram each broken line stands for a factor $-
\left(4 \pi \ell^2_0\right)^{d/2} \beta_e$, and a propagator line of
momentum ${\bf k}$ connecting special points $j_\alpha < j_\beta$
stands for $\exp \left(-{\bf k}^2 \ell^2_0 \left(j_\beta - j_\alpha
  \right) \right)$. Internal momenta are integrated over all space:
\begin{eqnarray*}
\int \frac{d^d k}{\left(2 \pi \right)^d} \dots \equiv \int\limits_{{\bf k}} \dots \: \: \: ,
\end{eqnarray*}
and the segment labels of the special points are summed from $1$ to
$n-1$, respecting their ordering along the chain.

With these rules we can evaluate ${\cal Z}\left({\bf q},0 \right)$.
The derivative $\nabla_h |_0 \,{\cal Z} \left({\bf q}, {\bf h} \right)$
involves a further vertex, drawn as a stroke through segment $j$ in
the polymer line. It arises from the differentiation $\nabla_h |_0$
and yields a factor $2 i \ell^2_0 {\bf k}$, with ${\bf k}$ being the
internal momentum flowing into that special point. The segment index
$j$ is not summed over. An explicit example for the application of
these rules is given below.

\begin{figure}
\includegraphics[scale=0.5]{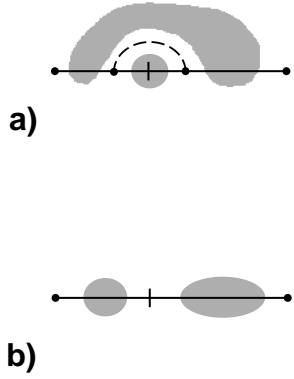}%
\caption{\label{fig:1}Classes of diagrams discussed in the text. The
  grey blobs stand for any number of interactions.}
\end{figure}

\begin{figure}
\includegraphics[scale=0.5]{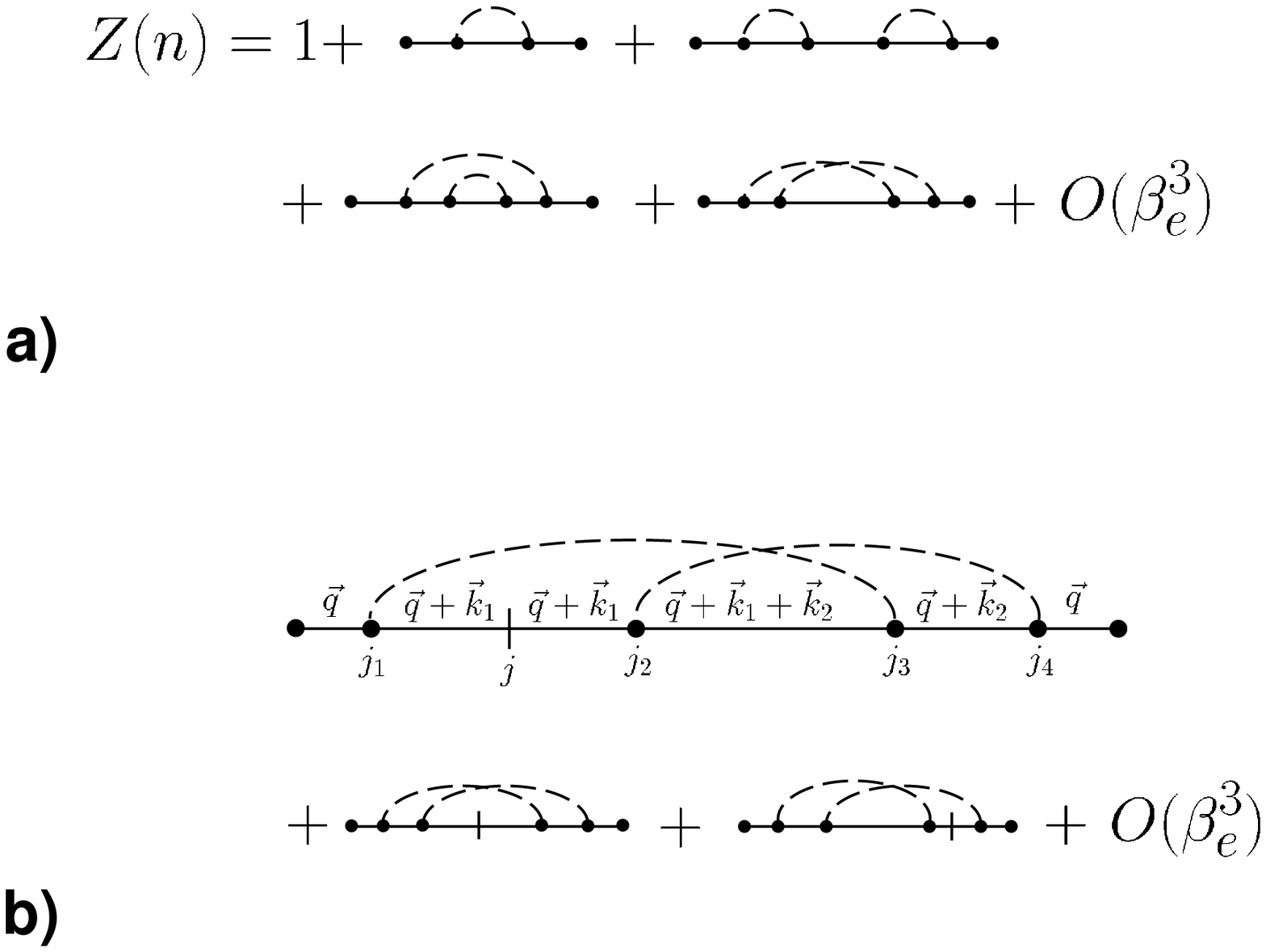}%
\caption{\label{fig:2}a) Diagrammatic expansion of $Z\left(n\right)$
  b) Diagrams contributing to $Z\left(n\right)
  \hat{L}^{\mbox{\scriptsize (irr)}}$}
\end{figure}

In evaluating $\hat{L}\left(j,n\right)$ (Eq.~2.8), the contribution of
each diagram of the general structure of Fig.~\ref{fig:1}a vanishes.
Here ${\bf s}_j$ is part of a polymer loop closed by the explicitly
drawn interaction. The loop does not interact with the remainder of
the chain and therefore the direction of ${\bf s}_j$ is not correlated
with ${\bf r}_n -{\bf r}_0$. Diagrams of the structure shown in
Fig.~\ref{fig:1}b yield a very simple contribution. Due to momentum
conservation the momentum flowing into the stroke is ${\bf q}$, and
under the operator $-i \,\nabla_q |_0$ the total contribution of all
such diagrams yields $ 2d \ell^2_0
\,Z\left(j\right)Z\left(n-j\right)$. We thus find a `reducible'
contribution
\begin{equation}
\hat{L}^{\left({\mbox{\scriptsize red}}\right)}\left(j,n\right) = 2 d
\ell^2_0\,\frac{Z\left(j\right) Z\left(n-j\right)}{Z\left(n\right)} \: \: \: .
\end{equation}
Up to order $\beta^2_e$ the diagrams for $Z(n)$ are shown in Fig.~\ref{fig:2}a.
The remaining `irreducible' contributions
$\hat{L}^{\mbox{\scriptsize (irr)}}\left(j,n\right)$ are at least of
order $\beta^2_e$, as shown in Fig.~\ref{fig:2}b. To exemplify the application
of the evaluation rules we write down the contribution to
$\hat{L}^{\mbox{\scriptsize (irr)}}\left(j,n\right)$ of the explicitly
labeled diagram:
\begin{eqnarray}
\left(4 \pi \ell^2_0\right)^d \beta^2_e 2 \ell^2_0 \nabla_q \bigg|_0 \,\sum^{j-1}_{j_1=1} \;\sum_{j<j_2<j_3<j_4<n} \; \int\limits_{{\bf k}_1} \int\limits_{{\bf k}_2} &&\nonumber  \\
e^{-{\bf q}^2 \ell^2_0 j_1} e^{-\left({\bf q} + {\bf k}_1\right)^2 \ell^2_0 \left(j-j_1\right)} \left({\bf q} + {\bf k}_1\right) e^{-\left({\bf q} + {\bf k}_1\right)^2 \ell^2_0 \left(j_2 -j\right)} &&\nonumber \\
e^{- \left({\bf q} + {\bf k}_1 + {\bf k}_2\right)^2 \ell^2_0 \left(j_3 - j_2\right)} 
e^{- \left({\bf q} + {\bf k}_2\right)^2 \ell^2_0 \left(j_4 - j_3\right)} 
e^{-{\bf q}^2 \ell^2_0 \left(n-j_4\right)}
 &=& 2 d \ell^2_0 \beta^2_e D_1\left(j\right)
\end{eqnarray}
\begin{equation}
D_1\left(j\right) = \sum^{j-1}_{j_1=1} \;\sum_{j<j_2<j_3<j_4<n} 
    \frac{\left(j_3-j_2\right) \left(j_4-j_3\right)}{\left[ \left(j_4 - j_2\right) \left(j_3-j_1\right) - \left(j_3-j_2\right)^2 \right]^{1+d/2}} \: \: \: .
\end{equation}
The second diagram of Fig.~\ref{fig:2}b yields $2d \ell^2_0 \beta^2_e D_2(j)$,
where
\begin{equation}
D_2\left(j\right) = - \sum_{0<j_1<j_2<j} \;\sum_{j<j_3<j_4<n} 
\frac{\left(j_2-j_1\right) \left(j_4-j_3\right)}{\left[ \left(j_4 - j_2\right) \left(j_3-j_1\right) - \left(j_3-j_2\right)^2 \right]^{1+d/2}} \: \: \: ,
\end{equation}
whereas the last diagram results from the first one by reflection of
the chain $\left(j \to n-j\right)$. We thus find
\begin{equation}
\hat{L}^{{\mbox{\scriptsize (irr)}}} \left(j,n\right) = 2 d \ell^2_0 \beta^2_e \left[ D_1\left(j\right) + D_2\left(j\right) + D_1 \left(n-j\right) \right] + {\cal O} \left(\beta^3_e \right) \: \: \: .
\end{equation}
Now it is clear that only the irreducible diagrams yield new
contributions specific for the persistence length. It is for this
reason that we evaluated $\hat{L}(j,n)$ including order $\beta^2_e$.

\section{\label{sec:3}Calculation of $\hat{L}\left( \lowercase{j,n}\right)$}
\subsection{First order unrenormalized perturbation theory}
In order $\beta_e$ only the first diagram in Fig.~\ref{fig:2}a contributes:
\begin{eqnarray}
Z\left(n\right) & = & 1 - \left(4 \pi \ell^2_0\right)^{d/2} \beta_e \sum_{0<j_1<j_2<n} \int\limits_{\bf k} e^{- {\bf k} {\ell^2_0} \left(j_2 - j_1 \right)} + {\cal O}\left(\beta^2_e\right) \nonumber \\
& = & 1 - \beta_e \sum_{0<j_1<j_2<n} \left(j_2 - j_1 \right)^{-d/2} + {\cal O}\left(\beta^2_e\right) \: \: \: .
\end{eqnarray}
Substituting this result into Eq.~(2.9) for
$\hat{L}^{{\mbox{\scriptsize red}}}\left(j,n\right)$ we find
\begin{eqnarray}
\hat{L}\left(j,n\right)  & = & 2 d \ell^2_0 \left[1 + \beta_e R_1\left(j\right) + {\cal O}\left(\beta^2_e\right) \right] \: \: \: , \\
R_1\left(j\right) & = & \sum^{j-1}_{j_1 =1} \;\sum^{n-1}_{j_2 = j + 1} \left(j_2 - j_1\right)^{- d/2} \: \: \: .
\end{eqnarray}
For $\epsilon = 4 - d > 0$ the summations in $R_1$ can be approximated
by integrals, which implies that we take the limit of a continuous
chain. This yields
\begin{equation}
\frac{\hat{L}\left(j,n\right)}{2 d \ell^2_0} = 1 + \beta_e n^{\epsilon/2} \frac{4}{\epsilon \left(2-\epsilon\right)} \left[\bar{\jmath}^{\; \epsilon/2} + \left(1-\bar{\jmath}\right)^{\epsilon/2} - 1\right] + {\cal O}\left(\beta^2_e\right) \: \: \: ,
\end{equation}
where we introduced the notation
\begin{equation}
\bar{\jmath} = \frac{j}{n} \: \: \: .
\end{equation}
The result (3.4) is correct for $n \gg 1$, up to terms of relative
order $1/n$ neglected in the continuous chain limit.

\subsection{Renormalization}
Renormalization exploits the fact that for $n \gg 1$ the
microstructure becomes unimportant and physical observables are
invariant under a change of $\ell_0$, compensated by an appropriate
change of $\beta_e$ and $n$. The theory has often been explained in
the literature, and we here use the formulation of Refs.~\cite{S7,S8}.

We introduce a renormalized length scale
\begin{equation}
\ell_R = \frac{\ell_0}{\lambda} \: \: \: ,
\end{equation}
and we define the renormalized coupling $u$ and the renormalized chain
length $n_R$ by the formal relations
\begin{eqnarray}
\beta_e & = & \lambda^\epsilon u Z_u\left(u\right) \: \: \: , \\
n & = & \lambda^{-2} n_R Z_n\left(u\right) \: \: \: .
\end{eqnarray}
The ratio $\bar{\jmath} = j/n$ is invariant under renormalization. The
parameter $\lambda$ obeys the inequality $0 < \lambda < 1$, but
otherwise is arbitrary. The renormalization factors $Z_u$ and $Z_n$
are chosen to absorb the poles in $\epsilon$, which show up in
expressions like Eq.~(3.4). In the continuous chain limit implicit in
the evaluation of segment summations as integrals, these poles carry
the information on the microstructure. Within the scheme of minimal
subtraction the renormalization factors up to the order needed here
are found as
\begin{eqnarray}
Z_u\left(n\right) & = & \frac{1}{2} \left(1 + \frac{4}{\epsilon} u + {\cal O} \left(u^2 \right) \right) \: \: \: , \\ 
Z_n\left(u\right) & = & 1 - \frac{u}{\epsilon} - \left( \frac{3}{2\epsilon^2} - \frac{5}{8 \epsilon} \right) u^2 + {\cal O} \left(u^3\right) \: \: \: .
\end{eqnarray}

The renormalization factors have been calculated to higher orders in
$u$, and very accurate expressions for the logarithmic derivatives
$\partial \ln u/ \partial \ln \lambda, \: \partial \ln Z_n/ \partial
\ln \lambda$ have been derived, known as renormalization group flow
equations (see e.g.~Ref.~\cite{S9}). For $\lambda \to 0$, which
corresponds to the excluded volume limit of long self-repelling
chains, the renormalized coupling tends to a fixed point $u^*$. With
our convention one finds
\begin{equation}
u^* = 0.364\:,\quad \left(d=3\right) \: \: \: .
\end{equation}
We also will need the $\epsilon$--expansion
\begin{equation}
u^* = \frac{\epsilon}{4} + \frac{21}{128} \epsilon^2 + {\cal O} \left(\epsilon^3\right) \: \: \: .
\end{equation}
Integrating the flow equations one derives the renormalization group
mapping from bare to renormalized parameters. In terms of a normalized
coupling
\begin{equation}
f = \frac{u}{u^*}
\end{equation}
it reads
\begin{eqnarray}
\ell_R & = & f |1-f|^{-1/\omega} H_u \left(f\right) s_\ell \: \: \: , \\
n_R & = & f^{-2} |1-f|^{1/\left(\nu \omega\right)} \frac{H\left(f\right)}{H^2_u\left(f\right)} s_n n \: \: \: .
\end{eqnarray}
Here $s_\ell = \ell_0 \,\bar{s}_\ell (\beta_e)$, $s_n = s_n
(\beta_e)$ are integration constants, which contain the
dependence on the bare parameters $\ell_0$, $\beta_e$. In practice they
are to be taken as microscopic fit parameters. $\omega$ and $\nu$ are
critical exponents, which in three dimensions take the values
\begin{equation}
\nu \approx 0.588\:; \quad \omega \approx 0.80 \: \: \: .
\end{equation}
We below also will need the $\epsilon$--expansion of $\nu$:
\begin{equation}
\nu =\frac{1}{2} + \frac{\epsilon}{16} + \frac{15}{512} \epsilon^2 + {\cal O}\left(\epsilon^3\right) \: \: \: .
\end{equation}
Finally, the functions $H_u\left(f\right)$ and $H \left(f\right)$
within the accuracy of the mapping can be parameterized as
\begin{eqnarray}
H_u\left(f\right) & = & \left(1 + 0.824 f \right)^{0.25} \: \: \: , \\
H\left(f\right) & = & 1 - 0.005 f - 0.028 f^2 + 0.022 f^3 \: \: \: .
\end{eqnarray}

So far we just recalled some general results of renormalization group
theory. We now turn to the renormalization of the persistence length.
The sum rule (1.3) indicates that $n \hat{L} \left(j,n\right) \sim
\langle \left({\bf r}_n -{\bf r}_0\right)^2 \rangle$. Since
$R^2_e \left(n\right) = \langle \left({\bf r}_n -{\bf r}_0\right)^2
\rangle$
is an observable, invariant under the renormalization group, this
suggests to define the renormalized persistence length as
\begin{equation}
L_R \left(\bar{\jmath},n_R\right) = \frac{n}{n_R} \hat{L}\left(j,n\right) = \lambda^{-2} Z_n \left(u\right) \hat{L}\left(j,n\right) \: \: \: .
\end{equation}
Indeed, using Eqs.~(3.4), (3.6) -- (3.10), we find
\begin{equation}
\frac{L_R\left( \bar{\jmath},n_R\right)}{2d \ell^2_R} = 1 - \frac{u}{\epsilon} + u n^{\epsilon/2}_R \frac{2}{\epsilon \left(2-\epsilon\right)} \left[\bar{\jmath}^{\; \epsilon/2} + \left(1 - \bar{\jmath}\right)^{\epsilon/2} -1\right] + {\cal O} \left(u^2\right)
\end{equation}
For $\epsilon \to 0$ this yields
\begin{equation}
\frac{L_R\left(\bar{\jmath},n_R\right)}{2d \ell^2_R} = 1 + \frac{u}{2} \Big[1 + \ln n_R + \ln \bar{\jmath} + \ln \left(1-\bar{\jmath}\right) + {\cal O} \left(\epsilon\right) \Big] + {\cal O}\left(u^2\right) \: \: \: .
\end{equation}
The pole in $\epsilon$ is cancelled, as expected for a properly
renormalized quantity.

\subsection{Crossover analysis of the first order result}
Though by construction the leading long chain behavior of observables
like $R^2_e$ or $n \hat{L}\left(j,n\right)$, if evaluated to all
orders of perturbation theory, is invariant under renormalization, low
order perturbative approximations generally depend on our choice of
the renormalized length scale $\ell_R$. An exception from this rule is
provided by so called `universal ratios', i.e. dimensionless ratios of
observables, constructed such that all explicit renormalization
factors drop out. Evaluated at the fixed point $u^*$, such ratios have
a unique $\epsilon$--expansion, independent of any conventions of the
renormalization scheme. In the next subsection we will construct such
a ratio. Here we are concerned with the direct evaluation of the
result (3.21), and we thus have to face the problem of the proper
choice of $\ell_R$.

Previous work on many different experimental observables shows
\cite{S8} that we can construct a good approximation by evaluating
first order results like Eq.~(3.21) directly in three dimensions
$\left(\epsilon = 1\right)$, with $\ell_R$ taken to be of the order of
the smallest length relevant to the problem considered. In other
words, we use the renormalization group to map the physical chain on a
chain of effective segments (`blobs') of size $\ell_R$, chosen such
that the quantity considered does not resolve the internal structure
of a blob. For the present problem the blob should be identified with
the smaller one of the two subchains $\left(0,j\right)$,
$\left(j,n\right)$. To construct a smooth crossover among the limits
$j \to 0$ and $j \to n$ we define a chain length variable
\begin{equation}
\hat{n} = \frac{j \left(n-j\right)}{n} = n \bar{\jmath} \left(1-\bar{\jmath}\right) \: \: \: ,
\end{equation}
and we implicitly fix $\ell_R = \ell_0/\lambda$ by choosing
\begin{equation}
\hat{n}_R = \lambda^2
Z^{-1}_{n} \hat{n} = n_0 \: \: \: ,
\end{equation}
where $n_0$ is a constant of order 1. To stay consistent with our
previous work \cite{S8} we take $n_0 = 0.53$, a value determined from
an analysis of the interpenetration ratio, which is proportional to
the second virial coefficient of the osmotic pressure divided by the
coil volume.

Taking $\epsilon = 1$ we find from Eqs.~(3.20), (3.21), (3.23), (3.24)
\begin{eqnarray}
\hat{L} \left(j,n\right) & = & \frac{\hat{n}_R}{\hat{n}}\, L_R \left(\bar{\jmath} ,n_R\right) \nonumber \\ 
& = & 6\, \frac{n_0}{\hat{n}} \,\ell^2_R \left[1 + 2 u^* f n^{1/2}_0 \left( \left(1 - \bar{\jmath} \right)^{-1/2} + \bar{\jmath}^{\; -1/2} - \bar{\jmath}^{\; -1/2}\left(1-\bar{\jmath}\right)^{-1/2} - \frac{1}{2} \,n^{-1/2}_0 \right)\right] \: \: \: .
\end{eqnarray}
Eqs.~(3.14), (3.15) yield
\begin{eqnarray}
n_0 & = & f^{-2}|1-f|^{\frac{1}{\nu \omega}} \frac{H\left(f\right)}{H^2_u\left(f\right)} s_n \hat{n} \: \: \: , \\
\frac{n_0}{\hat{n}} \ell^2_R & = & |1-f|^{\frac{1}{\omega}\left(\frac{1}{\nu}-2 \right)} H\left(f\right) s^2_\ell s_n \: \: \: .
\end{eqnarray}
As a result we find the crossover form of $\hat{L} \left(j,n\right)$,
evaluated to first order renormalized perturbation theory:
\begin{eqnarray}
\hat{L}\left(j,n\right) & = & 6 s_n s^2_\ell |1-f|^{\frac{1}{\omega}\left(\frac{1}{\nu}-2 \right)} H\left(f\right) \nonumber \\ 
& & \cdot\left[1 + 2u^* f n^{1/2}_0 \left(\left(1-\bar{\jmath}\right)^{-1/2} + \bar{\jmath}^{\; -1/2} - \bar{\jmath}^{\; -1/2} \left(1-\bar{\jmath}\right)^{-1/2} - \frac{1}{2} \,n^{-1/2}_0 \right) \right] \: \: \: .
\end{eqnarray}
This result deserves a detailed discussion.

Eq.~(3.28) involves the parameter $s_\ell$, which has dimensions of a
length, only in the combination
\begin{equation}
\tilde{\ell}^2 = s_n s^2_\ell \: \: \: .
\end{equation}
This is a general feature of the renormalized theory. At the
$\Theta$--point, which in the present theory corresponds to a strictly
noninteracting chain $\left(\beta_e = 0 = f \right)$, $\tilde{\ell}$
reduces to $\ell_0$. (For $T \gtrsim \Theta$ it is weakly temperature
dependent: $\tilde{\ell} = \ell_0 \left(1 + {\cal
    O}\left(\beta_e\right)\right)$, $\beta_e \sim T - \Theta$.)
Evaluating Eq.~(3.28) at the $\Theta$--point $f = 0$ we thus find the
expected result:
\begin{equation}
\hat{L}\left(j,n\right) = 6 \ell^2_0 = \langle {\bf s}^2_j \rangle , \quad T = \Theta \: \: \: . 
\end{equation}
The persistence length is of microscopic size, independent of $n$ and
$j$. Note that keeping residual three body interactions and treating
the $\Theta$--point as a tricritical point we in three dimensions
expect to find logarithmic corrections similar to those found for the
end--to--end distance \cite{S10}: $\hat{L}\left(j,n\right) \approx 6
\ell^2_0 \left(1 + \mbox{\small const}/ \ln n\right)$.

The excluded volume limit $\beta_e > 0$, $n \to \infty$, is reached for
$f \to 1$. Using Eq.~(3.26) to eliminate $|1-f|$ we from Eq.~(3.28)
find
\begin{eqnarray}
\hat{L}^* \left(j,n\right) & = & 6 B^2 n_0 \left(n \bar{\jmath} \left(1-\bar{\jmath}\right) \right)^{2\nu-1} \nonumber \\ 
& & \cdot\left[ 1 + 2u^* n^{1/2}_0 
\left(\left(1-\bar{\jmath}\right)^{-1/2} + \bar{\jmath}^{\; -1/2} - \bar{\jmath}^{\; -1/2} \left(1-\bar{\jmath}\right)^{-1/2} - \frac{1}{2} \,n^{-1/2}_0 \right) \right] \: \: \: ,
\end{eqnarray}
where, as usual, the star indicates the excluded volume limit. The
microscopic length parameter $B$ is defined as
\begin{equation}
B = s_\ell s^\nu_n n^{-\nu}_0 H^{1-2\nu}_u(1) H^\nu (1) \: \: \: . 
\end{equation}
It is the only microscopic parameter showing up in the excluded volume
limit.

The result (3.31) shows the expected overall scaling $\hat{L}^* \sim
n^{2\nu-1}$. It furthermore predicts some pronounced dependence of
$\hat{L}^*\left(j,n\right)$ on $\bar{\jmath}$. For $\bar{\jmath} \ll
1$ it yields
\begin{eqnarray}
\hat{L}^*\left(j,n\right) & = & 6 B^2 n_0 \left(n \bar{\jmath}\right)^{2\nu-1} \left[1 + u^* \left(2 n^{1/2}_0 - 1 + {\cal O} \left(\bar{\jmath}^{1/2}\right) \right) \right] \nonumber \\ 
& \sim & j^{0.176} \: \: \: .
\end{eqnarray}
Thus the persistence length is microscopic close to the chain ends and
rapidly increases for $\bar{\jmath}$ approaching the center of the
chain. We also note that $\hat{L}^*\left(j,n\right)$ for $j \ll n$ (or
$n-j \ll n$) is independent of the chain length. This feature is quite
plausible. It implies that the blob $(0,j)$ essentially is influenced
only by its neighboring blobs along the chain. Starting from this
hypothesis we easily can derive the basic structure of Eq.~(3.31) from
a simple scaling argument. Assuming that $\hat{L}\left(j,n\right)$ is
renormalizable, we from the sum rule (1.3) find the scaling law
\begin{equation}
\hat{L}^*\left(j,n\right) = n^{2\nu-1} \hat{\cal L}\left(\bar{\jmath}\right) \: \: \: .
\end{equation}
Then assuming that the limits $n \to \infty$, $j$ fixed, and $n \to
\infty$, $(n-j)$ fixed, exist, we immediately find
\begin{equation}
\hat{\cal L}\left(\bar{\jmath}\right) = \left(\bar{\jmath}\left(1-\bar{\jmath}\right) \right)^{2\nu-1} {\cal L}_1\left(\bar{\jmath}\right) \: \: \: ,
\end{equation}
where ${\cal L}_1\left(0\right) = {\cal L}_1\left(1\right)$ takes some
finite value.

The variation of $\hat{L}^*\left(j,n\right)$ (Eq.~(3.31)) closely
resembles the behavior of the end--to--end swelling factor
$\alpha^2_E\left(\hat{n}\right)$ of a chain of length $\hat{n} = n
\bar{\jmath} \left(1-\bar{\jmath}\right)$. The latter is defined as
\begin{equation}
\alpha^2_E \left(\hat{n}\right) = \frac{R^2_e \left(\hat{n}\right)}{6 \tilde{\ell}^2 \hat{n}} \: \: \: ,
\end{equation}
where the denominator is the mean squared end--to--end distance of a
noninteracting reference chain, reducing to the physical chain only at
the $\Theta$--point. (Recall the remark below Eq.~(3.29).) The close
similarity among $\hat{L}\left(j,n\right)$ and $\alpha^2_E
\left(\hat{n}\right)$ holds also outside the excluded volume limit.
Within our renormalization scheme we to first order renormalized
perturbation theory find
\begin{equation}
\alpha^2_E\left(\hat{n}\right) = |1-f|^{\frac{1}{\omega}\left(\frac{1}{\nu}-2\right)} H\left(f\right) \left[1 + u^* f n^{1/2}_0 \left(\frac{2}{3} - n_0^{-1/2} \right)\right] \: \: \: .
\end{equation}
From Eqs.~(3.28), (3.29), (3.37) we may construct the ratio
\begin{equation}
\frac{\hat{L}\left(j,n\right)}{6 \tilde{\ell}^2 \alpha^2_E \left(\hat{n}\right)} = 
\frac{1 + 2u^* f n^{1/2}_0 \left( \left(1-\bar{\jmath}\right)^{-1/2} + \bar{\jmath}^{\; -1/2} - \bar{\jmath}^{\; -1/2} \left(1-\bar{\jmath}\right)^{-1/2} - \frac{1}{2}\, n^{-1/2}_0\right)}
{1 + u^* f n^{1/2}_0 \left(\frac{2}{3} - n^{-1/2}_0 \right)} \: \: \: . 
\end{equation}
Here the prefactor
$|1-f|^{\frac{1}{\omega}\left(\frac{1}{\nu}-2\right)}$, which contains
the dominant variation, has dropped out, leaving only some weak
dependence on $\bar{\jmath}$ and the coupling strength $f$. For
$\bar{\jmath} = \frac{1}{2}$ the ratio (3.38) varies from 1 for $f =
0$ to about 1.3 for $f=1$. We thus find $\hat{L}\left(j,n\right) \sim
\alpha^2_E \left(\hat{n}\right)$ quite generally. Indeed, if evaluated
at the fixed point $f=1$,
\begin{equation}
\rho = \frac{\hat{L}\left(j,n\right)}{6 \tilde{\ell}^2 \alpha^2_E \left(\hat{n}\right)} 
\equiv \frac{\hat{n} \hat{L}\left(j,n\right)}{R^2_e \left(\hat{n}\right)}
\end{equation}
is an universal ratio, which will be calculated to order $\epsilon^2$
in the next subsection. We there also will compare the crossover of
$\hat{L}\left(j,n\right)$ from $\Theta$--conditions $(f=0)$ to the
excluded volume limit $(f=1)$, as resulting from the present analysis,
to the result of second order $\epsilon$--expansion (see Fig.~\ref{fig:4}).

\subsection{Expansion to order $\epsilon^2$}
In the analysis of the previous subsection we implicitly have built in
the power law behavior (3.35)
\begin{eqnarray*}
\hat{L}^*\left(j,n\right) \sim \left(n \bar{\jmath} \left(1-\bar{\jmath}\right) \right)^{2\nu-1} \sim j^{2\nu-1}\:,\quad (\bar{\jmath} \ll 1)\:\:\:,
\end{eqnarray*}
by choosing the renormalized length scale $\ell_R$ to be of the order
of the end--to--end distance of the smaller subchain. This amounts to
incorporating the blob picture underlying the scaling approach, and is
adequate only if the limit $n \to \infty$, $j$ fixed, is
finite. The validity of that approach is not evident a priory. Other observables related to the internal structure of an excluded volume chain, like the distribution of distances among segments $j_1$ and $j_2$, are known to show specific end effects, showing different power laws close to a chain end as compared to the center of the chain. It therefore is conceivable that also $L(j,n)$ shows new singularities for $\bar{\jmath} \to 0$. In Sect.~\ref{sec:4B} we will show this to be the case in two dimensions. Here we check the behavior of $L^*(j,n)$ within the framework of strict $\epsilon$--expansion. The logarithmic terms showing up in this expansion must sum up to the expected power law.

It is easily checked that the first order result (3.22) obeys this
criterion. Using the $\epsilon$--expansions of $u^*$ (Eq.~(3.12)) and
$\nu$ (Eq.~(3.17)) we find
\begin{eqnarray}
\frac{\hat{L}_R\left(\bar{\jmath},n_R\right)}{2 d \ell^2_R} & = & 1 + \frac{\epsilon}{8} \left(1 + \ln n_R + \ln \bar{\jmath} + \ln \left(1 - \bar{\jmath}\right)\right) + {\cal O} \left(\epsilon^2\right) \nonumber \\ 
& = & \left(1 + \frac{\epsilon}{8}\right) \left(n_R \,\bar{\jmath}\left(1-\bar{\jmath}\right)\right)^{\epsilon/8} + {\cal O}\left(\epsilon^2\right) \nonumber \\ 
& = & \left(1 + \frac{\epsilon}{8}\right) \left(n_R \,\bar{\jmath}\left(1-\bar{\jmath}\right)\right)^{2\nu-1} + {\cal O}\left(\epsilon^2\right) \: \: \: .
\end{eqnarray}
However, as pointed out at the end of Sect.~\ref{sec:2}, the irreducible
contributions specific to $\hat{L}\left(j,n\right)$ occur first in
second order. It thus is appropriate to calculate
$L_R\left(\bar{\jmath},n\right)$ to order $\epsilon^2$.

Unrenormalized expressions for the irreducible diagrams have been
given in Sect.~\ref{sec:2}, Eqs.~(2.10) -- (2.11). For the reducible
contribution we find
\begin{eqnarray}
\frac{\hat{L}^{\mbox{\scriptsize (red)}} \left(j,n\right)}{2 d \ell^2_0} & = & 1 + \beta_e R_1 \left(j\right) + \beta^2_e \left(R^2_1 \left(j\right) + R_2\left(j\right) + R_2\left(n-j\right) \right. \nonumber \\ 
& & + \left. R_3\left(j\right) + R_3\left(n-j\right) - R_4\left(j\right) \right) + {\cal O}\left(\beta^3_e\right) \: \: \: ,
\end{eqnarray}
where $R_1\left(j\right)$ has been defined in Eq.~(3.3), and
\begin{eqnarray}
R_2\left(j\right) & = & \sum^{j-1}_{j_1=1} \;\sum_{j<j_2<j_3<j_4<n}
\left(j_3 - j_2\right)^{-d/2} \left[ \left(j_4 - j_1\right)^{-d/2} -
  \left(j_4 - j_3 + j_2 - j_1\right)^{-d/2} \right] \: \: \:,
\\
R_3\left(j\right) & = & \sum^{j-1}_{j_1=1} \;\sum_{j<j_2<j_3<j_4<n}
\bigg[ \left( j_3 - j_1\right)^{-d/2} \left(j_4 - j_2\right)^{-d/2} 
\nonumber\\
&&\quad\quad\quad\quad\quad\quad\quad
-\left( \left(j_3 - j_1\right) \left(j_4-j_2\right) -
    \left(j_3-j_2\right)^2 \right)^{-d/2} \bigg] \: \: \:,
\\
R_4\left(j\right) & = & \sum_{0 < j_1 < j_2 < j} \;\sum_{j < j_3 < j_4 < n} \bigg[ \left(j_3 - j_2\right)^{-d/2} \left(j_4 - j_3 + j_2 - j_1\right)^{-d/2} \nonumber  \\ 
& & \quad\quad\quad\quad\quad\quad\quad\quad + \left( \left(j_3 - j_1\right) \left(j_4 - j_2\right) - \left(j_3-j_2 \right)^2\right)^{-d/2} \bigg] \: \: \:. 
\end{eqnarray}
Here the diagrammatic contributions have been combined such that for
$\epsilon > 0$ all summations can be evaluated as integrals. The
evaluation of expressions (3.42) -- (3.44), (2.11), (2.12) is
straightforward, but lengthy.  The resulting unrenormalized expansion
reads
\begin{equation}
\frac{\hat{L}\left(j,n\right)}{2 d \ell^2_0} = 1 + \beta_e n^{\epsilon/2} a_1\left(\bar{\jmath}\right) + \beta^2_e n^{\epsilon} a_2\left(\bar{\jmath}\right) + {\cal O} \left(\beta^3_e\right) \: \: \: ,
\end{equation}
with coefficients given in $\epsilon$--expansion as
\begin{eqnarray}
a_1\left(\bar{\jmath}\right) & = & \frac{2}{\epsilon} + 1 + \ln\bar{\jmath} + \ln \left(1-\bar{\jmath}\right) \nonumber \\ 
& & + \frac{\epsilon}{2} \left(1 + \ln\bar{\jmath} + \ln \left(1-\bar{\jmath}\right) + \frac{1}{2} \ln^2 \bar{\jmath} + \frac{1}{2} \ln^2 \left(1-\bar\jmath\right) \right) + {\cal O}\left(\epsilon^2\right) \: \: \: , 
\\[2ex]
a_2\left(\bar{\jmath}\right) & = & - \frac{6}{\epsilon^2} - \frac{1}{\epsilon} \left( \frac{13}{2} + 6 \ln \bar{\jmath} + 6 \ln \left(1 - \bar{\jmath}\right)      \right) \nonumber \\ 
& & - \frac{61}{8} + \frac{\pi^2}{3} - \frac{13}{2} \ln \bar{\jmath} - \frac{13}{2} \ln \left(1 - \bar{\jmath}\right) - 3 \ln^2 \bar{\jmath} - 3 \ln^2 \left(1 - \bar{\jmath}\right) \nonumber \\ 
& & + \ln \bar{\jmath} \: \ln \left(1 - \bar{\jmath}\right) + {\cal F} \left(\bar{\jmath}\right) + {\cal F} \left(1 - \bar{\jmath}\right) + {\cal O} \left(\epsilon\right) \: \: \: ,
\end{eqnarray}
with
\begin{eqnarray}
{\cal F}\left(\bar{\jmath}\right) & = & \frac{1}{4} \ln \bar{\jmath} \: \ln \left(1 - \bar{\jmath}\right) - \sqrt{\bar{\jmath}} \: \sqrt{4 - 3 \bar{\jmath}} \: \ln \frac{\sqrt{4 - 3 \bar{\jmath}} + \sqrt{\bar{\jmath}}}{2 \sqrt{1 - \bar{\jmath}}} \nonumber \\
& & - \ln \sqrt{1 - \bar{\jmath}} + \left(7 - 9 \bar{\jmath} + 3 \bar{\jmath}^2\right) \ln^2 \frac{\sqrt{4 - 3 \bar{\jmath}} + \sqrt{\bar{\jmath}}}{2 \sqrt{1 - \bar{\jmath}}} - \ln^2 \sqrt{1 - \bar{\jmath}} \nonumber \\
& & + \int\limits^{\bar{\jmath}}_{0} dt \left( 9 - 6 t\right) \ln^2 \frac{\sqrt{4 - 3 t} + \sqrt{t}}{2 \sqrt{1 - t}} \: \: \: .
\end{eqnarray}
The function ${\cal F}\left(\bar{\jmath}\right)$ contains no leading
logarithmic terms, but vanishes for $\bar{\jmath} \to 0$ and stays
finite for $\bar{\jmath} \to 1$.  We now use Eqs.~(3.6) -- (3.10),
(3.20) to find the renormalized expression
\begin{eqnarray}
\frac{L_R \left(\bar{\jmath}, n_R\right)}{2 d \ell^2_R}  =  1&+& \frac{u}{2} \left[ 1 + \ln \hat{n}_R + \frac{\epsilon}{2} \left( 1 + \ln \hat{n}_R + \frac{1}{2} \ln^2 \hat{n}_R - \ln \bar{\jmath} \: \ln \left(1 - \bar{\jmath} \right) \right) + {\cal O} \left(\epsilon^2\right) \right] \nonumber \\
&+& \frac{u^2}{4} \left[  \frac{\pi^2}{3} - \frac{45}{8} - \frac{9}{2} \ln \hat{n}_R - \frac{3}{2} \ln^2 \hat{n}_R  + 4 \ln \bar{\jmath} \: \ln\left(1-\bar{\jmath} \right) \right. \nonumber \\
&+&{\cal F}\left(\bar{\jmath}\right) + {\cal F} \left(1 - \bar{\jmath}\right) + {\cal O} \left(\epsilon\right) \bigg] + {\cal O} \left(u^3\right) \: \: \: , 
\end{eqnarray}
where
\begin{equation}
\hat{n}_R = n_R \bar{\jmath} \left(1 - \bar{\jmath}\right) \: \: \: ,
\end{equation} 
as above. The absence of any $\epsilon$--poles in Eq.~(3.49) verifies
the renormalizability of the persistence length to second order in
$u$.

To check the power law (3.35) we put $n_R = 1$, and we evaluate
\begin{equation}
\bar{\cal L} \left(\bar{\jmath}\right) = \left(\bar{\jmath} \left(1 -
    \bar{\jmath}\right)\right)^{1-2 \nu}
\frac{L_R\left(\bar{\jmath},1\right)}{2 d \ell^2_R}
\end{equation}
for $u = u^*$ in strict $\epsilon$--expansion, using Eqs.~(3.12),
(3.17). Some algebra yields
\begin{equation}
\bar{\cal L} \left(\bar{\jmath}\right) = 1 + \frac{\epsilon}{8} + \frac{\epsilon^2}{64} \left[ \frac{\pi^2}{3} + \frac{29}{8} + {\cal F}\left(\bar{\jmath}\right) + {\cal F} \left(1-\bar{\jmath}\right) \right] + {\cal O} \left(\epsilon^3\right) \: \: \: . 
\end{equation}
All the leading logarithmic singularities are eliminated, the
remaining singular terms contained in ${\cal
  F}\left(\bar{\jmath}\right)$ and ${\cal F}\left(1 -
  \bar{\jmath}\right)$ being of order $\bar{\jmath} \, \ln
\bar{\jmath}$ and $\left(1 - \bar{\jmath}\right) \ln \left(1 -
  \bar{\jmath}\right)$. They thus only give rise to subleading corrections to the dominant behavior $L(j,n) \sim j^{2 \nu -1}$, $j \ll n$. This proves the asymptotic power law to second
order in $\epsilon$.

For a quantitative evaluation or our result we consider the ratio
(3.39):
\begin{eqnarray*}
\rho\left(j,n\right) = \frac{\hat{L}\left(j,n\right)}{2 d \tilde{\ell}^2 \alpha^2_E \left(\hat{n}\right)} \: \: \: .
\end{eqnarray*}
The second order result for $\alpha^2_E \left(\hat{n}\right)$ reads
\begin{equation}
\frac{\hat{n}\tilde{\ell}^2}{\hat{n}_R \ell^2_R} \alpha^2_E \left(\hat{n}\right) = \alpha^2_{E,R} \left(\hat{n}_R\right)
\end{equation}
\begin{eqnarray}
\alpha^2_{E,R} \left(\hat{n}_R\right)  =  1 &-& \frac{u}{2} \left(1 - \frac{\epsilon}{2} - \ln \hat{n}_R \left(1 - \frac{\epsilon}{2}\right) - \frac{\epsilon}{4} \ln^2 \hat{n}_R + {\cal O} \left(\epsilon^2\right) \right) \nonumber \\
&-& \frac{u^2}{4} \left(\frac{5}{8} - \frac{\pi^2}{12} + \frac{7}{3} \alpha_0 - \frac{3}{2} \ln \hat{n}_R + \frac{3}{2} \ln^2 \hat{n}_R + {\cal O} \left(\epsilon\right) \right) +\mathcal{O}\left(u^3\right)\: \: \: ,
\end{eqnarray}
where
\begin{equation}
\alpha_0 = \int\limits^{1}_{0} dt \,\frac{\ln t}{1 - t + t^2} \: \: \: .
\end{equation} 
With relation (3.20) among $\hat{L}\left(j,n\right)$ and $L_R
\left(\bar{\jmath},n_R\right)$ we find
\begin{equation}
\rho\left(j,n\right) = \frac{L_R \left(\bar{\jmath},n_R \right)}{2 d \ell^2_R \alpha^2_{E,R} \left(\hat{n}_R\right)} \: \: \: .
\end{equation}
We now use Eqs.~(3.49), (3.54) together with
\begin{eqnarray*}
u = u^* f = \frac{\epsilon}{4} \left( 1 + \frac{21}{32} \,\epsilon \right) f \: \: \: ,
\end{eqnarray*}
to find in strict $\epsilon$--expansion
\begin{eqnarray}
\rho\left(j,n\right) & = & 1 + \frac{\epsilon}{4} f + \frac{21}{128}
\, \epsilon^2 f + \frac{\epsilon^2}{16}  f \left(1 - f\right) \left(
  2 \ln \hat{n}_R - \ln \bar{\jmath} \: \ln \left(1 -
    \bar{\jmath}\right) \right) \nonumber \\
& & + \frac{\epsilon^2}{64} f^2 \left(\frac{\pi^2}{4} - 3 +
  \frac{7}{3} \, \alpha_0 + {\cal F}\left(\bar{\jmath}\right) + {\cal
    F} \left(1 - \bar{\jmath}\right) \right) + {\cal O}
\left(\epsilon^3 \right) \: \: \: .
\end{eqnarray}

We first consider the fixed point $f = 1$, where $\rho
\left(j,n\right)$ reduces to
\begin{equation}
\rho^*\left(\bar{\jmath}\right) = 1 + \frac{\epsilon}{4} +
\frac{\epsilon^2}{6 4} \left(\frac{\pi^2}{4} + \frac{15}{2} +
  \frac{7}{3} \, \alpha_0 + {\cal F}\left(\bar{\jmath}\right) + {\cal
    F}\left(1 - \bar{\jmath}\right) \right) + {\cal O}
\left(\epsilon^3\right) \: \: \:.
\end{equation}
We note that the dependence on $\ln \hat{n}_R$ and thus on the choice
of the renormalized length scale has dropped out, as expected for a
critical ratio at the excluded volume fixed point.

To check the result (3.58), we may exploit the sum rule (1.3):
\begin{eqnarray*}
R^2_e \left(n\right) = \sum^{n}_{j = 1} \hat{L}\left(j,n\right) \: \: \: .
\end{eqnarray*}
By virtue of the power law ${R^2_e}^* \left(\hat{n}\right) =
\left(\bar{\jmath} \left(1 - \bar{\jmath}\right) \right)^{2\nu}
{R_e^2}^* \left(n\right)$, which holds in the excluded volume limit,
the sum rule immediately takes the form
\begin{equation}
\int\limits^{1}_{0} d \bar{\jmath} \left(\bar{\jmath} \left(1 - \bar{\jmath}\right) \right)^{2 \nu-1} \rho^*\left(\bar{\jmath}\right) = 1 \: \: \: .
\end{equation}
Employing the $\epsilon$--expansion (3.17) of $\nu$ and writing
\begin{eqnarray}
\rho^* \left(\bar{\jmath}\right) = 1 + \frac{\epsilon}{4} + \epsilon^2 \rho_2 \left(\bar{\jmath}\right) + {\cal O} \left(\epsilon^3\right) \: \: \: ,
\end{eqnarray}
we find from Eq.~(3.59)
\begin{equation}
\int\limits^{1}_{0} d \bar{\jmath} \: \rho_2 \left(\bar{\jmath}\right) = \frac{15}{128} + \frac{\pi^2}{384} \approx 0.1429 \: \: \: .
\end{equation}
We have checked by numerical integration that the result (3.58) obeys
this relation.

Despite its fairly complicated analytical form, (c.f. Eq.~(3.48)), the
function $\rho_2\left(\bar{\jmath}\right)$ is almost independent of
$\bar{\jmath}$. It monotonically decreases from $\rho_2 \left(0\right)
\equiv \rho_2 \left(1\right) \approx 0.1496$ to $\rho_2
\left(1/2\right) \approx 0.1405$. As a result, to the order considered
$\rho^*\left(\bar{\jmath}\right)$ essentially can be taken as
constant:
\begin{equation}
\rho^*\left(\bar{\jmath}\right) = 1.25 + \rho_2 \left(\bar{\jmath}\right) \approx \rho^* = 1.393 \: , \quad \left(\epsilon = 1\right) \: \: \: .
\end{equation}
If $\rho^*\left(\bar{\jmath}\right)$ were strictly independent of
$\bar{\jmath}$ to all orders of $\epsilon$, the sum rule (3.59) would
yield the result
\begin{equation}
\rho^* = \frac{\Gamma \left(4\nu\right)}{\Gamma^2 \left(2\nu\right)} \approx 1.408 \: , \quad \left(\epsilon = 1\right) \: \: \: .
\end{equation}
This is very close to the result of our second order calculation. The
small difference, left for the higher orders, suggests that also in
higher orders the $\bar{\jmath}$--dependence of $\rho^*
\left(\bar{\jmath}\right)$ stays weak.

\begin{figure}
\includegraphics{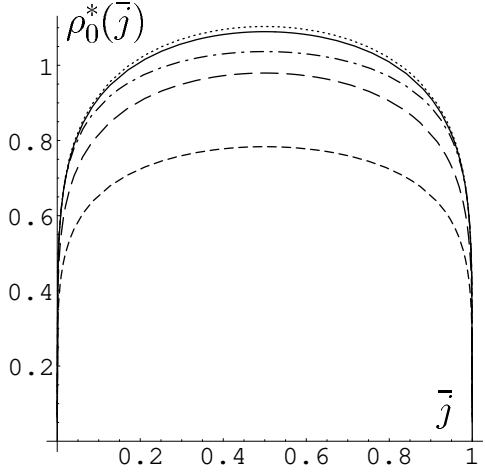}%
\caption{\label{fig:3}$\rho^*_0 \left(\bar{\jmath}\right)$
  (Eq.~(3.64)) as function of $\bar{\jmath}$. Results of the
  $\epsilon$--expansion: ${\cal O}\left(\epsilon^0\right)$, short
  dashes; ${\cal O}\left(\epsilon\right)$, long dashes; ${\cal
    O}\left(\epsilon^2\right)$, full line. One loop crossover model:
  dot--dashed line. Sum rule: dotted line.}
\end{figure}

To illustrate the $\bar{\jmath}$--dependence of $\hat{L}^*
\left(j,n\right)$ we in Fig.~\ref{fig:3} have plotted the universal ratio
\begin{equation}
\rho^*_0 \left(\bar{\jmath}\right) = \frac{n \hat{L} \left(j,n\right)}{R^{2^*}_e \left(n\right)} = \left(j \left(1 - \bar{\jmath}\right) \right)^{2\nu-1} \rho^*\left(\bar{\jmath}\right) \: \: \: ,
\end{equation}
evaluated in the different approximations discussed in this section.
Obviously the ${\cal O} \left(\epsilon^2\right)$--result (full line)
is almost indistinguishable from the hypothetical result (dotted
line) based on the sum rule $\rho^*\left(\bar{\jmath}\right) \equiv
\rho^* \approx 1.408$. The one loop crossover model (Eq.~(3.31),
dot-dashed line) coincides with the ${\cal O}
\left(\epsilon^2\right)$--result in the tails, but deviates somewhat
towards the center of the chain. This is understandable, since the
choice of the reference chain length $\hat{n}$ and thus of the
renormalized length scale is optimized for $\bar{\jmath} \to 0$ or
$\bar{\jmath} \to 1$. However, also in the center of the chain the one
loop crossover model considerably improves the plain ${\cal O}
\left(\epsilon\right)$--result (long dashed line).

So far we considered universal ratios. Turning to the normalized
persistence length $L\left(j,n\right)$, as defined in Eq.~(1.1), we
find from Eq.~(3.64)
\begin{equation}
L^*\left(j,n\right) = \frac{\hat{L}^*\left(j,n\right)}{\sqrt{\langle
    {\bf s}^2_j\rangle}} = \rho^*_0\left(\bar{\jmath}\right)
\frac{{R^2_e}^*\left(n\right)}{\sqrt{\langle {\bf s}^2_j\rangle} n} 
\: \: \: .
\end{equation}
For the discrete chain model bare perturbation theory shows that
$\langle {\bf s}^2_j \rangle$ for $d = 3$ weakly depends on $j$. This
is an endeffect, saturating for $j \gg 1$. Furthermore, taking the
continuous chain limit we find
\begin{eqnarray*}
\frac{\langle{\bf s}^2_j\rangle}{2 d \ell^2_0} \to f_s
\left(\beta_e\right)\:, \quad (d > 2)\:\:\:,
\end{eqnarray*}
where the function $f_s$ is independent of $\bar{\jmath}$, provided $0
< \bar{\jmath} < 1$. This suggests to introduce the arclength of the
chain
\begin{equation}
L_c = \sqrt{\langle {\bf s}^2_j \rangle} n \: \: \: ,
\end{equation}
neglecting any $j$--dependence. Employing the result (3.65) with
$\rho^*\left(\bar{\jmath}\right)$ taken from the sum rule (Eq.~3.63),
we thus find the simple expression
\begin{equation}
L^*\left(j,n\right) \approx 1.408 \left(\bar{\jmath}\left(1 -
    \bar{\jmath}\right) \right)^{2 \nu -1} \frac{R^{2*}_e
  \left(n\right)}{L_c}\:, \quad (d = 3) \: \: \: ,
\end{equation}
which generalizes a standard result \cite{S1} based on the wormlike
chain model. It, however, must be noted that in the present context
$\langle {\bf s}^2_j\rangle$ and thus $L_c$ are effective nonuniversal
quantities, so that $L_c$ may differ from the arclength calculated
from a physically realistic model by a factor of order 1.

We finally consider the crossover from $\Theta$--conditions to the
excluded volume limit. We use the representation
\begin{equation}
\frac{\hat{L}\left(j,n\right)}{2 d \tilde{\ell}^2} = \rho\left(j,n\right) \alpha^2_E \left(\hat{n}\right) \: \: \: ,
\end{equation}
where $\alpha^2_E \left(\hat{n}\right)$ is taken from Eq.~(3.53):
\begin{eqnarray*}
\alpha^2_E \left(\hat{n}\right) = \frac{\hat{n}_R \ell^2_R}{\hat{n}
  \tilde{\ell}^2} \,\alpha^2_{E,R} \left(\hat{n}_R\right) \: \: \: .
\end{eqnarray*}
Eqs.~(3.14), (3.15), (3.29) yield
\begin{equation}
\frac{\hat{n}_R \ell^2_R}{\hat{n} \tilde{\ell}^2} = |1 - f|^{\frac{1}{\omega} \left(\frac{1}{\nu} - 2\right)} H \left(f\right) \: \: \: .
\end{equation}
With the simple choice $\hat{n}_R = 1$ we find in strict
$\epsilon$--expansion of $\alpha^2_{E,R}$:
\begin{eqnarray}
  \alpha^2_E \left(\hat{n}\right) &= &|1 - f|^{\frac{1}{\omega}
    \left(\frac{1}{v} - 2\right)} H \left(f\right) \nonumber\\
  &&\cdot\left\{ 1 - \frac{\epsilon}{8} \left(1 + \frac{5}{32}\right) f
    -\frac{\epsilon^2}{64} f^2 \left( \frac{5}{8} - \frac{\pi^2}{12} +
      \frac{7}{3} \,\alpha_0 \right) + {\cal O}
    \left(\epsilon^3\right) \right\} \: \: \: .
\end{eqnarray}

\begin{figure}
\includegraphics{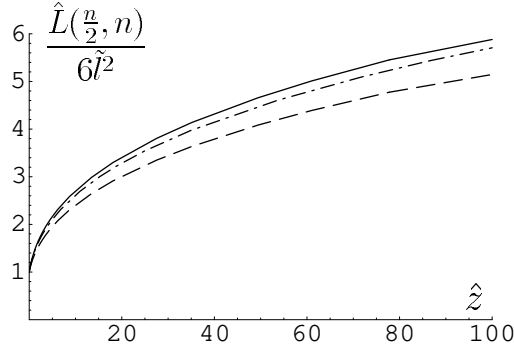}%
\caption{\label{fig:4}$\hat{L}\left(\frac{n}{2}, n\right)/6
  \tilde{\ell}^2$ as function of $\hat{z} = \sqrt{s_n \hat{n}}$. Full
  line: ${\cal O}\left(\epsilon^2\right)$; dashes: ${\cal
    O}\left(\epsilon\right)$; dot-dashed line: one loop crossover
  model.}
\end{figure}

In Fig.~\ref{fig:4} we plot the result for $\hat{L} \left(j = \frac{n}{2},
  n\right)/ 2 d \tilde{\ell}^2$ as function of $\hat{z} = \sqrt{s_n
  \left(\beta_e\right) \hat{n}}$, a variable which is the counterpart
in the renormalized theory of the standard $z$--variable of
two--parameter theory \cite{S1}. It is related to the intermediate
variable $f$ via Eq.~(3.15):
\begin{equation}
1 = f^{-2} |1 - f|^{1/\nu \omega}
\frac{H\left(f\right)}{H^2_u\left(f\right)} \,\hat{z}^2 \: \: \: .
\end{equation}
For a comparison we also show the result of the one-loop crossover
model (Eqs.~(3.38), (3.37), $u^* = 0.364, \: n_0 = 0.53$). (We recall,
that the choice $\hat{n}_R = n_0 = 0.53$ is optimized to the one loop
crossover model. No such analysis is available for the ${\cal O}
\left(\epsilon^2\right)$--calculation, so that we used $\hat{n}_R = 1$
as simplest possible choice.) Fig.~\ref{fig:4} illustrates the gradual increase
of $\hat{L}\left(j,n\right)$ with increasing excluded volume strength.
It furthermore again shows that the one-loop crossover model
considerably improves the plain ${\cal
  O}\left(\epsilon\right)$--result.

\section{\label{sec:4}Simulations}
We measured the persistence length in the Domb-Joyce model, where the
chain configuration is modelled as a random walk on a regular lattice.
Each configuration is weighted by a factor $\left(1-w\right)^{n_2}/Z$,
where $n_2$ is the number of pairwise intersections, and $Z =
Z\left(n\right)$ is the partition function. Using cubic or square
lattices, we can identify a microscopic segment with a primitive
lattice vector, which defines the unit of length so that
$\hat{L}\left(j,n\right) \equiv L\left(j,n\right)$. To simulate the
model we used the PERM--algorithm developed by Grassberger \cite{S11},
in the form also employed in our previous work \cite{S7} on the
correlations among segment directions.

\subsection{Results for $d=3$}
On the cubic lattice our simulations extend to a maximal chain length
$n_{\max} = 2000$. Most measurements were performed for $w = 0.4$,
which for this lattice is known to be close to the excluded volume
value $w^*$, where the leading corrections to scaling vanish, ($0.4 <
w^* < 0.5$, according to Ref.~\cite{S12}). In renormalized variables
$w^*$ corresponds to $f = \frac{u}{u^*} = 1$. To study the approach to
the excluded volume limit we also used $w = 0.01$ (weak coupling: $f
< 1$) and $w = 1$ (self-avoiding walks, strong coupling: $f > 1$).

\begin{figure}
\includegraphics{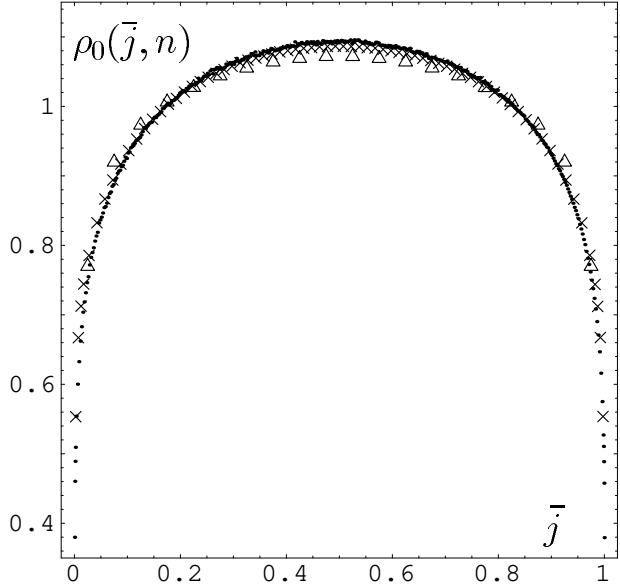}%
\caption{\label{fig:5}Simulation results for
  $\rho_0\left(\bar{\jmath},n\right) = n\,
  \hat{L}\left(j,n\right)/R^2_e\left(n\right)$ as function of
  $\bar{\jmath}$. Triangles: $n = 20$; crosses: $n = 200$; points: $n
  = 2000$.}
\end{figure}

Results for the ratio $\rho_0\left(\bar{\jmath},n\right)$ (Eq.~(3.64))
with $w = 0.4$ are shown in Fig.~\ref{fig:5}. We note that the results are
essentially independent of $n$, as expected for an universal ratio at
the fixed point. Only for the shortest chain $(n = 20)$ some small
deviation from scaling can be identified in this plot. These indicate
nonuniversal $1/n$--corrections. From the experimental side these
results establish the scaling law.

\begin{figure}
\includegraphics{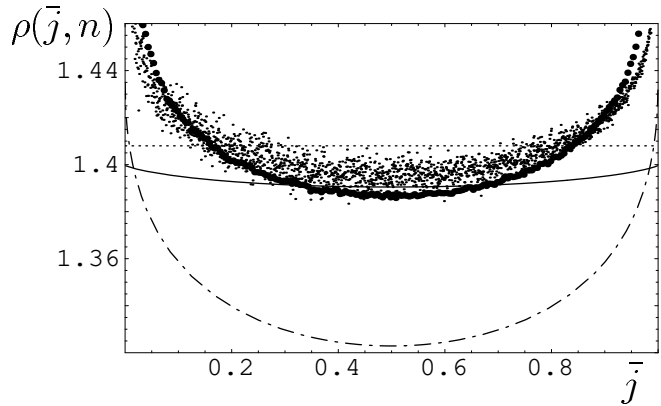}%
\caption{\label{fig:6}$\rho_0\left(\bar{\jmath},n\right) = \hat{n}\,
  \hat{L}\left(j,n\right)/R^2_e\left(\hat{n}\right)$ as function of
  $\bar{\jmath}$. Data: $n = 200$, large points; $n = 2000$, small
  points. Theory evaluated at the fixed point; full line: ${\cal
    O}\left(\epsilon^2\right)$; dot-dashed line: one loop crossover
  model; dotted line: sum rule (3.63).}
\end{figure}

In Fig.~\ref{fig:5} we have not included theoretical curves, since on
the scale of that figure the ${\cal O}\left(\epsilon^2\right)$--result
would just be covered by the data. For a comparison among theory and
data we therefore magnified the plot by dividing out the power law
$\bar{\jmath}\left(1 - \bar{\jmath}\right)^{2 \nu -1}$, resulting in
the ratio $\rho^*\left(\bar{\jmath}\right)$, (Eq.~3.39). To reduce the
data we used the relation $R^2_e\left(\hat{n}\right)/\hat{n} = \left(
  \bar{\jmath}\left(1 - \bar{\jmath}\right)\right)^{2 \nu - 1} R^2_e
\left(n\right)/n$, valid in the excluded volume limit.
Fig.~\ref{fig:6} shows the results for $n = 200$ and $n=2000$. We note
some small effect of chain length $n$: the data for $n = 2000$ seem to
trace out a flatter curve, being lifted in the center and lowered in
the wings compared to the data for $n = 200$. This is consistent with
$w = 0.4$ being slightly below the fixed point value $w^*$. For $0.2
\lesssim \bar{\jmath} \lesssim 0.8$ the data essentially fall between the
${\cal O} \left(\epsilon^2\right)$ result and the prediction of the
sum rule.  The increasing deviation among theory and data outside that
range just indicates that reducing the data by dividing out the power
law $\left(\bar{\jmath}\left(1-\bar{\jmath}\right)\right)^{2 \nu -1}$
strictly is adequate only at the fixed point and only for $j \gg 1$
and $n - j \gg 1$. With this in mind, and taking into account the
scale of the figure, we may state excellent agreement among theory and
data.  For completeness we note that the scatter in the data for $n =
2000$ decreases with increasing $j$. This is a feature of the
PERM--algorithm, which within a single numerical experiment executes
many correlated measurements. For the present problem the strength of
the correlation depends on $j$. A fair impression of the statistical
scatter is given by the data for small $j$.

\begin{figure}
\includegraphics{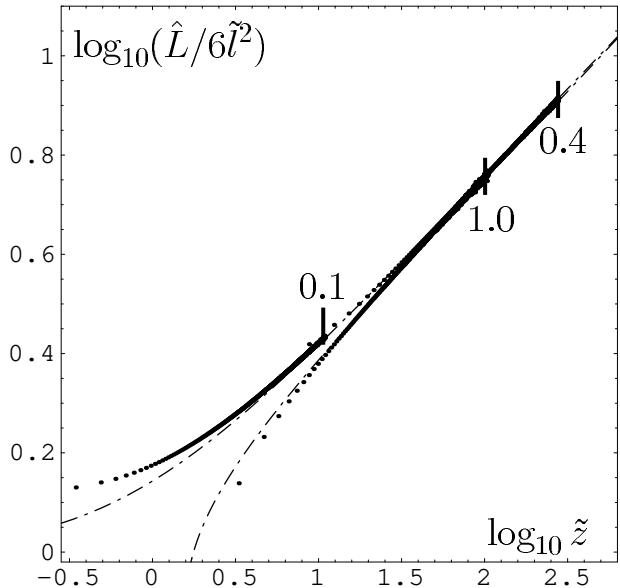}%
\caption{\label{fig:7}$\log_{10} \left(\hat{L}\left(\frac{n}{2},
      n\right)/ 6 \tilde{\ell}^2\right)$ as function of
  $\log_{10}\left(\tilde{v} n^{1/2}\right) = \log_{10} \tilde{z}$,
  measured for $w = 0.1$, $0.4$, $1.0$. Dot-dashed lines: one loop
  crossover model. The bars indicate $n_{\max} = 2000$ for the
  different $w$--values.}
\end{figure}

We finally consider the crossover towards excluded volume conditions.
Fig.~\ref{fig:7} shows a doubly logarithmic plot of $\hat{L}
\left(n/2,n\right)/6 \tilde{\ell}^2$ as function of $\tilde{z} =
\sqrt{s_n n}$. The nonuniversal parameters $\tilde{\ell}$ and
$\tilde{v} = \sqrt{s_n}$ for the Monte Carlo model as used here have
been determined in previous work \cite{S12} by fitting quantities like
the end--to--end distance to the one--loop crossover model. For
consistency we therefore also here compare the data to the one--loop
model (Eqs.~(3.28), (3.71)), noting that the difference to the full
${\cal O}\left(\epsilon^2\right)$ result is quite small, (see
Fig.~\ref{fig:4}).  The theoretical curves show the well known two
branched structure \cite{S12,S8} of the crossover scaling functions.
The upper branch represents crossover towards $\Theta$--conditions
$\left(f < 1\right)$, the lower branch corresponds to $f > 1$. This
structure is nicely confirmed by the data. As mentioned above, in
reducing the data we used the same parameter values $\tilde{\ell}$ and
$\tilde{v}$ as in Ref.~\cite{S12}, and we allowed for an additional
nonuniversal factor $\sim \sqrt{\langle{\bf s}^2_j\rangle}$, as
discussed in the context of Eq.~(3.67): independent of $w$ we
multiplied the data by $1.26$. We should note that in principle we
also should allow for corrections due to the discrete microstructure
of the Monte Carlo chain. This is particularly relevant for weak
excluded volume $\left(w = 0.1\right)$, where the measured persistence
length even for $n = 2000$ exceeds the microscopic bond length only by
a factor of about 2. Also the prefactor $\sim \sqrt{\langle{\bf s}^2_j
  \rangle}$ in principle should depend on $w$. Playing with such
corrections we obviously can bring the data in the range of small
$\tilde{z}$ even closer to the theory.  We, however, have not pursued
this further, being content with the good overall agreement among
theory and data exhibited in Fig.~\ref{fig:7}.

\subsection{\label{sec:4B}Some consideration of $d = 2$}
Grassberger \cite{S4}, and later Redner and Privman\cite{S5,S6},
considered the persistence length on two dimensional lattices, using
exact enumeration and simulations. They only discussed
$L\left(1,n\right)$, projecting the end--to--end vector on the
direction of the first segment. They found that $L\left(1,n\right)$
slowly increases with $n$, pointing to logarithmic behavior \cite{S5}:
$L \left(1,n\right) \sim \ln n$ or to a power law with a very small
exponent \cite{S4}: $L\left(1,n\right) \sim n^{0.063}$. In any case
this is quite different from the behavior in three dimensions, where
$L\left(1,n\right)$ stays microscopically small: $L\left(1,n\right)
\sim \sqrt{\langle s^2_1 \rangle}$, both according to our theory and
to our and previous \cite{S6} simulations.

\begin{figure}
\includegraphics{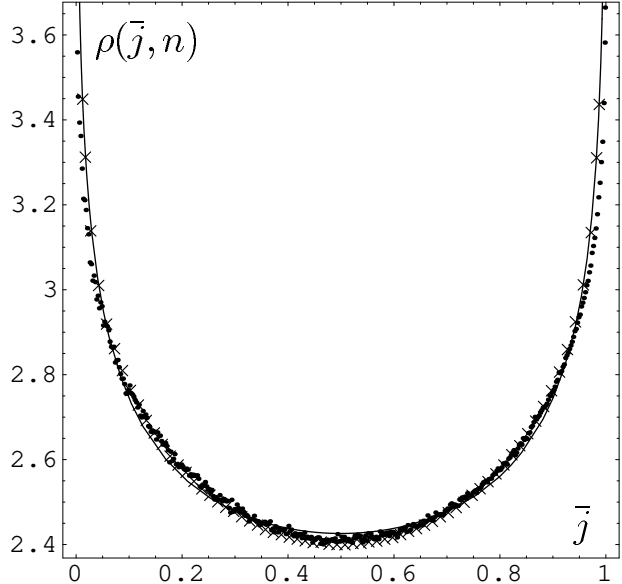}%
\caption{\label{fig:8}$\rho \left(\bar{\jmath},n\right)$ as function
  of $\bar{\jmath}$, measured on the square lattice. Data: crosses, $n
  = 200$; points, $n = 2000$. Excluded volume parameter: $w =
  0.4$. The curve gives a fit to Eq.~(4.2).}
\end{figure}

For a closer examination of this effect we carried through simulations
on the square lattice, using $w= 0.4$ and again going up to $n_{\max}
= 2000$. Results for the ratio
\begin{equation}
\rho\left(\bar{\jmath},n\right) = \left(\bar{\jmath}\left(1 - \bar{\jmath}\right) \right)^{- 1/2} \frac{n \hat{L}\left(j,n\right)}{R^2_e \left(n\right)}
\end{equation}
are shown in Fig.~\ref{fig:8}. (Note that $2\nu - 1 = 1/2$ in $d =
2$.) We again note a reasonable collapse of the data on a single
master curve. Note that the fixed point value $w^*$ is unknown for $d
= 2$, and $w = 0.4$ need not be close to $w^*$. Thus the data collapse
indicates that the chain lengths $n \gtrsim 200$ in $d = 2$ are
sufficient to reach the excluded volume limit. We thus believe that
our results indicate that the scaling law (3.34):
\begin{eqnarray*}
\hat{L}^*\left(j,n\right) = n^{2\nu-1} \hat{\cal L}\left(\bar{\jmath}\right) 
\end{eqnarray*}
also holds in two dimensions. However, the scaling function evidently
does not obey the simple law
\begin{eqnarray*}
\hat{\cal L}\left(\bar{\jmath}\right) \sim \left(\bar{\jmath} \left(1 - \bar{\jmath}\right)\right)^{2\nu-1} \: \: \: .
\end{eqnarray*}

The difference to three dimensions is drastically illustrated by
comparing Figs.~6 and 8, which both up to a constant give $\hat{\cal
  L}\left(\bar{\jmath}\right) / \left(\bar{\jmath}\left(1 -
    \bar{\jmath}\right)\right)^{2\nu-1}$. Indeed, in two dimensions
the data are reasonably well fitted by the form
\begin{equation}
\rho\left(j,n\right) = a \left[ 1 - b \left(\frac{\ln \left(1 - \bar{\jmath}\right)}{\bar{\jmath}} + \frac{\ln \bar{\jmath}}{1 - \bar{\jmath}} \right) \right] \: \: \: ,
\end{equation}
which for $\bar{\jmath} = 1/n$ yields logarithmic behavior
\begin{eqnarray*}
\hat{L}\left(1,n\right) \sim \ln n \: \: \: ,
\end{eqnarray*}
consistent with previous findings. This is illustrated by the full
curve in Fig.~\ref{fig:8}, which gives the expression (4.2) with $b$ taken as
fitparameter. $a = a\left(b\right)$ was determined from the sum rule
(3.59). The resulting parameter values are found as $b = 0.275$ and $a
= 1.376$.

Is there any support for the existence of such logarithmic terms from
the side of the theory? It first must be noted that the
$\epsilon$--expansion clearly does not yield such terms, but the naive
extrapolation from $d = 4$ down to $d = 2$ is somewhat doubtful.
Indeed, evaluating the unrenormalized first order contribution (3.2),
(3.3) directly for $d = 2$, we find a leading behavior
\begin{equation}
\beta_e R_1\left(j\right) \approx - \beta_e \hat{n} \left( \frac{\ln \left(1-\bar{\jmath}\right)}{\bar{\jmath}} + \frac{\ln \bar{\jmath}}{1 - \bar{\jmath}}\right) \: \: \: .
\end{equation}
(It is this result, which motivated the ansatz (4.2) for fitting
$\hat{\cal L} \left(\bar{\jmath}\right)$.) This logarithmic anomaly is
specific to two dimensions and cannot be found in the
$\epsilon$--expansion. In the context of polymer physics such
anomalies first have been discussed by Des Cloizeaux. (See chapters
12, Sect.~3.2.4 and 10, Sect.~4.2.6 of Ref.~\cite{S13}.) He found that
such anomalies occur in dimensions $d = 4 - 2/p$ ($p$ integer). For
the partition function $Z\left(n\right)$ they yield a prefactor, which
depends both on the usual variable $z = \beta_e n^{\epsilon/2}$ of the
unrenormalized two parameter model and on $\ln n$. In standard
observables like $R^2_e$ this prefactor cancels and no anomalies show
up. The results of Refs.~\cite{S4,S5,S6} as well as those presented
here suggest, that for the persistence length in $d = 2$ the anomaly
survives in the form of terms depending on $\ln \left(j/n\right), \:
\ln \left(\left(n - j\right)/n\right)$. This is not inconsistent with
des Cloizeaux's findings, which only imply the absence of $\ln
n$--terms. Indeed, exploiting Eq.~(12.3.49) of Ref.\cite{S13} we
immediately find that the reducible contribution
$\hat{L}^{\mbox{\scriptsize (red)}} \left(j,n\right)$
(Eq.~(2.9)) in bare perturbation theory for $d = 2$ picks up an
anomalous prefactor
\begin{eqnarray*}
g \left(\beta_e n, \bar{\jmath}\right) \exp \left[ -\, C_1 \beta_e \hat{n} \left(\frac{\ln \left(1 - \bar{\jmath}\right)}{\bar{\jmath}} + \frac{\ln \bar{\jmath}}{1 - \bar{\jmath}} \right)\right]
\end{eqnarray*}
where the constant $C_1$ and the amplitude function $g$ can be
calculated perturbatively. Note that the combination of the anomalous
terms is the same as in Eqs.~(4.2) and (4.3), our fit formula (4.2) just
taking into account the lowest order correction following from des
Cloizeaux's work.

For the irreducible contribution the same anomaly occurs. This is a
necessary prerequisite for renormalizability, which mixes $L^{\mbox
  {\scriptsize (red)}} \left(j,n\right)$ and $L^{\mbox {\scriptsize
    (irr)}} \left(j,n\right)$. We, however, have not pursued the
matter further, since a short calculation shows that also the
normalizing factor $\sqrt{\langle{\bf s}^2_j\rangle}$ of $L
\left(j,n\right)$ (Eq.~(1.1)) is plagued by anomalies: $\langle {\bf
  s}^2_j \rangle = 4 \ell^2_0(1 + \mbox {const} \,\beta_e
\ln\bar{\jmath})$ for $n \to \infty$, $d = 2$.  This sheds some doubt
on the very applicability of the self-repelling Gaussian chain model
for a calculation of the persistence length in two dimensions. (We
recall that $\langle {\bf s}^2_j \rangle$ in three dimensions rapidly
tends to a constant.) Irrespective of this concern this discussion
shows that also theoretical arguments support the existence of a
logarithmic anomaly in two dimensions.

\section{\label{sec:5}Conclusions}
As expected, we have found that the persistence length of an excluded
volume chain is a critical quantity, which can be evaluated within the
framework of renormalized two parameter theory. We have proven the
renormalizability to two loop order. Furthermore, considering the
excluded volume limit of a long self repelling chain we both in theory
and in simulations have found that the persistence length in three
dimensions is well approximated by the surprisingly simple expression
\begin{eqnarray*}
L^*\left(j,n\right) \approx 1.408 \left( \frac{j}{n} \left(1 - \frac{j}{n}\right) \right)^{0.176} \frac{{R^2_e}^* \left(n\right)}{L_c} \: \: \: ,
\end{eqnarray*}
where $L_c$ is the effective contour length of the chain. This expression contains both the overall scaling $L^* \sim n^{2\nu -1}$ and the dominant variation along the chain. The full ${\cal O}\left(\epsilon^2\right)$--result predicts some additional complicated dependence on $\bar{\jmath} = j/n$, but numerically this is completely irrelevant. Our result shows that in $d = 3$ the persistence length is of microscopic size for $j$ close to a chain
end, consistent with previous exact enumerations \cite{S6}. It rapidly
increases towards the center of the chain. This is consistent with our
previous results \cite{S7} on the correlations among segment
directions, which show a similar end--effect. It may be of interest to
give an example of the size of $L^* \left(j,n\right)$ for a typical
system. For $j = n/2$ and $n = 2000$ our simulations yield
$L^*\left(n/2,n\right) \approx 3.36$, to be compared to ${R_e}^*
\left(n\right) \approx 78$, both measured in units of the lattice
spacing, which here is identical to Kuhn's effective segment length.
Since the chain length in the Monte Carlo model roughly corresponds to
the polymerization index of a highly flexible polymer, this gives an
impression of the typical persistence length in a medium size polymer
coil.

Outside the excluded volume limit our results show that
$L\left(j,n\right)$ essentially varies like the end--to--end swelling
factor of a chain of effective length $\hat{n} = j \left(1 -
  j/n\right)$, with a prefactor which slowly increases from 1
($\Theta$--point) to about 1.4 (excluded volume limit). Again this
result is in good accord with our simulations.

For two--dimensional systems both our simulations and theoretical
arguments point to the existence of a logarithmic anomaly:
$L\left(j,n\right) \sim \ln n$ ($n \to \infty$, $j$ fixed),
thus supporting previous findings \cite{S4,S5,S6}. However, a precise
analytical analysis might need a model with segments of fixed length,
not the Gaussian chain model employed here.

With respect to the much more delicate \cite{S2} problem of an effective, locally defined persistence length in polyelectrolytes our results might be used to eliminate the excluded volume effects, if the persistence length is determined by fitting the observed coil radius to a worm--like chain model. This is a strategy sometimes followed in the analysis of data, (see e.g. Refs.~\cite{S14,S15}). Our results also point to a strong variation of the persistence length along the chain.


\begin{acknowledgments}
  We want to thank J. Hager for numerous discussions and support in
  the early stage of the simulations. This work has been supported by
  the Deutsche Forschungsgemeinschaft, SFB 237.
\end{acknowledgments}


\begin{thebibliography}{88}
\bibitem{S1} H.~Yamakawa, \textit{Modern Theory of Polymer Solutions}, Harper
  \& Row, New York, 1971
\bibitem{S2} M.~Ullner, C.~E.~Woodward, Macromolecules {\bf 35} (2002)
  1437
\bibitem{S3} R.~Everaers, A.~Milchev, V.~Yamakov, Eur.~Phys.~J. E {\bf
    8} (2002) 3
\bibitem{S3b} R.~R.~Netz, H.~Orland, Eur.~Phys.~J. {\bf 8} (1999) 81
\bibitem{S4} P.~Grassberger, Phys. Lett. {\bf 89 A} (1982) 381
\bibitem{S5} S.~Redner, V.~Privman, J.~Phys.~{\bf A}: Math. Gen. {\bf
    20}          (1987) L~857
\bibitem{S6} V.~Privman, S.~Redner, Z.~Phys.~{\bf B 67} (1987) 129
\bibitem{S7} L.~Sch\"afer, A.~Ostendorf, J.~Hager, J.~Phys.~{\bf A}:
  Math.  Gen. {\bf 32} (1999) 7875
\bibitem{S8} L.~Sch\"afer, \textit{Excluded Volume Effects in Polymer
    Solutions}, Springer, Heidelberg, 1999
\bibitem{S9} R.~Schloms, V.~Dohm, Nucl.~Phys.~{\bf B 328}(1989) 639
\bibitem{S10} B.~Duplantier, J.~Physique~{\bf 41} (1980) L~409
\bibitem{S11} P.~Grassberger, Phys.~Rev.~E~{\bf 56} (1997) 3682
\bibitem{S12} P.~Grassberger, P.~Sutter, L~Sch\"afer, J.~Phys.~{\bf
    A}: Math.~ Gen.~{\bf 30} (1997) 7039
\bibitem{S13} J.~des~Cloizeaux, G.~Jannink, \textit{Polymers in Solution},
  Clarendon Press, Oxford, 1990
\bibitem{S14} U.~Micka, K.~Kremer, Phys.~Rev.~{\bf E 54} (1996) 2653
\bibitem{S15} S.~F\"orster, M.~Schmidt, M.~Antonietti,
  J.~Phys.~Chem.~{\bf 96} (1992) 4008
\end{thebibliography}

\end{document}